\documentstyle[preprint,aps]{revtex}

\begin{document}

\draft

\title
{Gravitational Radiation from Nonaxisymmetric Instability in a
 Rotating Star}

\author{J. L. Houser, J. M. Centrella, and S. C. Smith}
\address
{Department of Physics and Atmospheric Science, Drexel University,
Philadelphia, PA  19104}

\maketitle

\begin{abstract}
We present the first calculations of the gravitational radiation
produced by nonaxisymmetric dynamical instability in a rapidly rotating
compact star. The star deforms into a
bar shape, shedding $\sim 4\%$ of its mass and $\sim 17\%$ of
its angular momentum.  The gravitational radiation is calculated in the
quadrupole approximation.  For a
 mass $M \sim 1.4$ M$_{\odot}$ and radius $R \sim 10$ km,
the gravitational waves have frequency $\sim 4$ kHz
and amplitude $h \sim 2 \times 10^{-22}$ at the distance
of the Virgo Cluster. They carry off energy
$\Delta E/M \sim 0.1\%$ and radiate angular momentum
$\Delta J/J \sim 0.7\%$.
\end{abstract}


\narrowtext

The detection of gravitational waves from astrophysical sources is one
 of the key frontiers for research in general relativity \cite{LIGO92}.
 Among the
 sources that might be observed by detectors currently under
development are those due to global rotational instabilities in
collapsing or compact stars \cite{thorne92&87,schutz86}.  For
 example, a rapidly rotating stellar core that has exhausted its
nuclear fuel and is prevented from collapsing to neutron star size by
centrifugal forces could become unstable, potentially shedding enough
 angular momentum to allow it to collapse and form a supernova
\cite{thorne93}.  In addition, neutron stars spun up by accretion of
 mass from a binary companion could possibly reach fast enough rotation
rates to go unstable \cite{schutz89,wagoner}.  The prospect of several
gravitational wave detectors becoming operational within a decade
means that detailed modeling of such sources and the radiation they
produce has a high priority.

Global rotational instabilities arise in fluids from growing modes
 $e^{\pm im\phi}$, where $m=2$ is the so-called ``bar mode''.  They
 are conveniently parametrized by $\beta \equiv T/|W|$, where $T$ is
the rotational kinetic energy and $W$ is the gravitational potential
energy; see \cite{schutz86,DT85,tassoul} for reviews.  We focus on the
bar instability since it is expected to be the fastest growing mode.
There are two different physical mechanisms by which this instability
can develop.  The dynamical bar instability is driven by Newtonian
 hydrodynamics and gravity.  It operates for fairly large values
 $\beta > \beta_d$ and develops on the timescale of about a rotation
period.  The secular instability is due to dissipative processes such
 as gravitational radiation reaction and occurs for typically lower
 values $\beta_s < \beta < \beta_d$.  It develops on a timescale of
several rotation periods or longer \cite{schutz89}.  For the uniformly
 rotating, incompressible, constant density Maclaurin spheroids
 $\beta_s \approx 0.14$ and $\beta_d \approx 0.27$.  Differentially
 rotating fluids with a polytropic equation of state
$P \propto \rho^{1 + 1/n}$, where $n$ is the polytropic index,
 are believed to undergo secular and dynamical bar instabilities
 at about these same values of $\beta$ \cite{DT85,ST}.

Astrophysical sources driven by rotational instabilities are
 nonlinear, time-dependent, and fully 3-dimensional systems;
 calculation of the gravitational radiation they produce requires
numerical modeling.  We have carried out computer simulations of
a differentially rotating compact star with a polytropic equation of
 state undergoing the dynamical bar instability.  This
 instability has previously been modeled numerically by Tohline
and collaborators in the context of star formation
\cite{DT85,TDM,DGTB,WT}.  Our work is the first to calculate
the gravitational radiation produced by this instability,
including waveforms and luminosities.
It is also a significant advance over the earlier studies because,
in addition to using
 better numerical techniques, we model the fluid correctly using
an energy equation.  This is essential due to the
generation of entropy by shocks during the later stages of the
evolution.

The gravitational radiation is calculated using Newtonian gravity
in the quadrupole approximation \cite{MTW}.  The gravitational
waveforms are given by the transverse-traceless (TT) components of
the metric perturbation $r\,h_{ij}^{\rm TT} = 2 \, {\skew6\ddot{
{I\mkern-6.8mu\raise0.3ex\hbox{-}}}}_{ij}\,^{\rm TT}$,
where ${{{I\mkern-6.8mu\raise0.3ex\hbox{-}}}}_{ij} =
\int\rho \,(x_i x_j -  {\textstyle{\frac{1}{3}}} \delta_{ij} r^2)
 \:d^3 r $ is the trace-reduced quadrupole moment of the source
and we use units in which $c = G = 1$.
Here, $r^2=x^2 + y^2 + z^2$, spatial indices $i,j=1,2,3$, and a dot
indicates a time derivative $d/dt$.  For an observer located on the
axis at $\theta=0, \phi=0$ in spherical coordinates centered on the
source, the waveforms take the simple form
$r{\sf h}^{\rm TT} = (\skew6\ddot{{I\mkern-6.8mu\raise0.3ex
\hbox{-}}}_{xx} - \skew6\ddot{{I\mkern-6.8mu\raise0.3ex\hbox
{-}}}_{yy}){\sf e}_{+} + 2 \skew6\ddot{{I\mkern-6.8mu\raise0.3ex
\hbox{-}}}_{xy}{\sf e}_
{\times}$,
where ${\sf e}_{+}$ and ${\sf e}_{\times}$ are basis tensors for
 the two polarization states \cite{KSTC}.
The gravitational wave luminosity is
$L = {\frac{1}{5}} \left\langle
{I\mkern-6.8mu\raise0.3ex\hbox{-}}^{(3)}_{ij}
{I\mkern-6.8mu\raise0.3ex\hbox{-}}^{(3)}_{ij}  \right \rangle$ and
 the angular momentum lost is $dJ_i/dt = {\frac{2}{5}} \epsilon_{ijk}
 \left\langle {I\mkern-6.8mu\raise0.3ex\hbox{-}}^{(2)}_{jm}
{I\mkern-6.8mu\raise0.3ex\hbox{-}}^{(3)}_{km}  \right \rangle$,
where the superscript $(3)$ indicates 3 time derivatives, there is
an implied sum on repeated indices, and the angle-brackets
indicate an average over several wave periods.  For these
burst sources such averaging is not well-defined; therefore we display
the unaveraged quantities ${\textstyle{\frac15}{I\mkern-6.8mu
\raise0.3ex\hbox{-}}^{(3)}_{ij}
{I\mkern-6.8mu\raise0.3ex\hbox{-}}^{(3)}_{ij}}$ and ${\textstyle
{\frac25}\epsilon_{ijk}{I\mkern-6.8mu\raise0.3ex\hbox{-}}^{(2)}_{jm}
{I\mkern-6.8mu\raise0.3ex\hbox{-}}^{(3)}_{km}}$ below.  The energy
emitted as gravitational radiation is $\Delta E = \int L\; dt$ and
the angular momentum carried away by the waves is $\Delta J_i =
\int (dJ_i/dt)dt$.

The equations of hydrodynamics are integrated using the techniques
of smoothed particle hydrodynamics (SPH) \cite{sph}, in which the
fluid is modeled as a collection of particles having non-zero extent
given by a smoothing length $h$.  The value of any physical field is
then obtained by averaging over all particles within $2h$ of a given
point using kernel estimation.   We have used the implementation of
SPH by Hernquist and Katz known as TREESPH \cite{HK}, which has
variable smoothing lengths and individual particle timesteps.
This code has
been well-tested by its developers; see \cite{HK} for details.
 To reduce the numerical noise
that can arise from taking derivatives of ${{I\mkern-6.8mu\raise0.3ex
\hbox{-}}}_{ij}$ numerically, we calculate $\skew6\ddot{{I\mkern-6.8mu
\raise0.3ex\hbox{-}}}_{ij}$ analytically from the SPH equations
of motion and obtain the gravitational waveforms directly from
quantities already available in the code (cf. \cite{FE});
this produces very smooth profiles.
The gravitational radiation calculation and the use of
 artificial viscosity to handle shocks have been
extensively tested and results are available in \cite{HK}
and \cite{CM}.

Our calculations begin with an initially axisymmetric differentially
rotating star with a polytropic equation of state.  This equilibrium
model is produced using the self-consistent field method of Smith
and Centrella \cite{SC}, which is based on earlier work of Ostriker
and Mark \cite{OM} and Hachisu \cite{hachisu86}.  This method is an
iterative technique carried out in cylindrical coordinates
$(\varpi, z, \phi)$.  A rotation law is specified by giving the
specific angular momentum as a function of mass interior to a
cylinder of radius $\varpi$.  Following the convention of earlier
work \cite{TDM,DGTB,WT,BO}, we use the rotation law for the rigidly
rotating, uniform density Maclaurin spheroids.  Since polytropes do
not have uniform density, this produces differentially rotating models.
The resulting density $\rho(\varpi,z)$ and angular velocity
$\Omega(\varpi)$ are then converted into a particle model to be
evolved with TREESPH.

The star has polytropic index $n = 3/2$, total mass $M$,
equatorial radius $R$, and $\beta = 0.30$.  This rapidly rotating
model is highly flattened, with polar radius $\sim 0.21 R$.  Time
is measured in units of the dynamical timescale for a sphere of
radius $R$,  $t_D \equiv (R^3 / M)^{1/2}$.
No perturbations are imposed upon the initial axisymmetric rotating
equilibrium model; the dynamical bar instability grows spontaneously
from small deviations from axisymmetry in the particle model. We
carried out runs with $N=2000, 4000, 8000$, and $16,000$
particles.  All models were run for $20 t_D$,
conserved total
energy to $\sim 1\%$ and angular momentum to $\sim 0.1\%$,
and produced generally similar results.
Table~\ref{table1} shows how some of the bulk properties of the
star vary with particle number.
We ran a set of tests to investigate
the effect of artificial viscosity on the growth of the bar
instability, comparing our results with the linearized tensor
virial analysis \cite{TDM}.  Since
the artificial viscosity used in ref. \cite{CM}
produced results that most closely matched the analytic
ones,
 we
used this form in the runs discussed here.
Both the growth rate and
rotation speed of the $m=2$ bar mode in all our runs
were closer to
the analytic results than the
previous numerical studies \cite{TDM}.

Fig.~\ref{xy-plots} shows the particle positions for the case
$N = 16,000$ projected onto the $x-y$ plane at 6 different times
during the calculation. The star is rotating counterclockwise about
the $z$ axis. The simulation starts at $t=0$, Fig.~\ref{xy-plots}(a).
  By $t=6.6 t_D$, Fig.~\ref{xy-plots}(b), nonaxisymmetric structure
is apparent.  As this bar shaped structure develops, the amplitude
of the $m=2$ mode grows exponentially until $t \sim 10 t_D$.  During
this time  a spiral arm pattern develops, with mass being shed from
the ends of the bar, as seen in Fig.~\ref{xy-plots}(c)
and~\ref{xy-plots}(d).  The bar and spiral arms exert gravitational
torques that cause angular momentum to be transported outward
\cite{DT85,TDM,DGTB,WT}.  The spiral arms expand supersonically
and merge together, causing shock heating and dissipation;
see Fig.~\ref{xy-plots}(e).  The system evolves toward a nearly
axisymmetric state with a core of equatorial radius $\sim R$
and $\beta \sim 0.26$, as shown in Fig.~\ref{xy-plots}(f);
c.f. \cite{RS}.
  The halo
contains $\sim 4\%$ of the total mass  and
$\sim 17\%$ of the angular momentum, and is bound.
   Throughout its evolution the entire system remains
flattened, with final polar radius $\sim 0.25 R$.

The gravitational waveforms $rh_+$ (solid line) and $rh_{\times}$
(dashed line) for this run are shown in Fig.~\ref{waves} for an
observer on the axis at $\theta = 0, \phi = 0$.
Fig.~\ref{lum} shows the gravitational wave luminosity $L$ and
Fig.~\ref{deltaE} shows the energy $\Delta E/M$ emitted as
gravitational waves.  The rate $dJ_z/dt$ at which angular momentum
is carried away by the gravitational waves is given in
Fig.~\ref{dJ/dt} and the angular momentum lost to gravitational
radiation $\Delta J_z/J$ is shown in Fig.~\ref{deltaJ}.
Scaling these results to a mass $M \sim 1.4$ M$_{\odot}$ and
radius $R \sim 10$ km, we find
that the gravitational radiation has frequency $f \sim 4$ kHz.
  The dimensionless amplitude of the waves is $h \sim 4
\times 10^{-19}$ at a distance $r \sim 10$ kpc, typical of
sources within the Milky Way, and $h \sim 2 \times 10^{-22}$ at a
distance of $r \sim 20$ Mpc, typical of sources in the Virgo
Cluster.  The energy lost to gravitational radiation is
$\Delta E/M \sim 0.1\%$ and the angular momentum radiated away is
$\Delta J/J \sim 0.7\%$.

In this Letter we present
 the first calculations of the gravitational
radiation produced by the dynamical bar instability.  This is an
important step in understanding astrophysical sources driven by
global rotational instabilities.  The dynamical instability may be
the mechanism by which a rapidly rotating, contracting stellar core
sheds enough angular momentum to allow collapse to neutron star
densities \cite{thorne-cutler}, and methods developed in this work
can be applied to study such collapses.  Neutron stars spun up by
accretion are expected to be subject to the secular instability
 \cite{schutz89}.
Modeling this process will require including the effects of
gravitational radiation reaction \cite{BDS} and
viscous fluids \cite{IL91}
in the hydrodynamical equations.

We thank L. Hernquist for supplying a copy of TREESPH and S. McMillan,
K. Thorne, and C. Cutler
for interesting and helpful discussions.
This work was supported by NSF grant
PHY-9208914 and the calculations were performed at the Pittsburgh
Supercomputing Center.

\begin{figure}
\caption{Particle positions are shown projected onto the $x-y$
plane for various times in the evolution of the model with
$N=16,000$ particles.}
\label{xy-plots}
\end{figure}

\begin{figure}
\caption{Gravitational waveforms $r h_+$ (solid line) and
$r h_{\times}$ (dashed line) are shown for the run with $N=16,000$.}
\label{waves}
\end{figure}

\begin{figure}
\caption{The gravitational wave luminosity $L\;(M/R)^{-5}$ is shown.
  This profile has been smoothed using simple averaging over a fixed
 interval of $0.1t_D$ centered on each point.}
\label{lum}
\end{figure}

\begin{figure}
\caption{The energy $[\Delta E/M ]\; (M/R)^{-7/2}$ emitted
in the form of gravitational
waves is given.}
\label{deltaE}
\end{figure}

\begin{figure}
\caption{The angular momentum $[dJ_z/dt]\; M^{-1} (M/R)^{-7/2}$
carried away by gravitational
 waves is shown.}
\label{dJ/dt}
\end{figure}

\begin{figure}
\caption{The total angular momentum $[\Delta J_z/J ]\; (M/R)^{-5/2}$
 lost to
gravitational radiation is shown.}
\label{deltaJ}
\end{figure}

\narrowtext
\begin{table}
\caption{The mass and angular momentum shed to the halo, and the
energy and angular momentum carried by the waves
(for $M = 1.4$ M$_{\odot}$ and $R = 10$ km), for
runs with different particle number.}
\begin{tabular}{ccccc}
 $N$ & $2000$ & $4000$ & $8000$ & $16,000$\\
\tableline
mass in halo & 6.0\% & 6.1\% & 5.7\% & 4.4\% \\
J in halo & 19\% & 19\% & 18\% & 17\% \\
$ \Delta E/M $ & .057\%  & .075\%  & .075\%  & .10\%  \\
$ \Delta J_z/J $ & .36\% & .52\%  & .52\%  & .72\%  \\
\end{tabular}
\label{table1}
\end{table}

\end{document}